\documentclass[journal]{IEEEtran}
\IEEEoverridecommandlockouts
\usepackage{mathrsfs}
\usepackage{svg}
\usepackage[ruled,linesnumbered]{algorithm2e}
\usepackage{verbatim}
\usepackage{soul}
\ifCLASSINFOpdf
\else
\fi
\usepackage[cmex10]{amsmath}
\usepackage{amsthm,cite,url,graphicx,booktabs,lipsum,color,bm,caption,subcaption,soul,makecell}
\usepackage{pifont,tikz,paralist,multirow,amssymb}
\theoremstyle{definition}

\makeatletter
\newcommand{\removelatexerror}{\let\@latex@error\@gobble}
\makeatother

\begin{document}
\title{Dynamic Client Clustering, Bandwidth Allocation, and Workload Optimization for Semi-synchronous Federated Learning}


\author{
Liangkun~Yu,~\IEEEmembership{Graduate~Student~Member,~IEEE,},
Xiang~Sun,~\IEEEmembership{Member,~IEEE,}
Rana~Albelaihi,
Chaeeun~Park,
and Sihua~Shao,~\IEEEmembership{Member,~IEEE}
\thanks{L. Yu, X. Sun, R. Albelaihi, and C. Park are with the Department of Electrical Engineering, University of New Mexico, Albuquerque, NM 87131, USA. E-mail: $\{$liangkun,sunxiang,ralbelaihi,chaeeun0618$\}$@unm.edu.\par
S. Shao is with the Department of Electrical Engineering, New Mexico Tech, Socorro, NM 87801, USA.\par
This work was supported by the National Science Foundation under Award under grant no. CNS-2323050 and CNS-2148178, where CNS-2148178 is supported in part by funds from federal agency and industry partners as specified in the Resilient \& Intelligent NextG Systems (RINGS) program.}
}
\markboth{ }%
{Shell \MakeLowercase{\textit{et al.}}: Bare Demo of IEEEtran.cls for IEEE Journals}
\maketitle

\begin{abstract}
Federated Learning (FL) revolutionizes collaborative machine learning among Internet of Things (IoT) devices by enabling them to train models collectively while preserving data privacy. FL algorithms fall into two primary categories: synchronous and asynchronous. While synchronous FL efficiently handles straggler devices, it can compromise convergence speed and model accuracy. In contrast, asynchronous FL allows all devices to participate but incurs high communication overhead and potential model staleness. To overcome these limitations, the semi-synchronous FL framework introduces client tiering based on computing and communication latencies. Clients in different tiers upload their local models at distinct frequencies, striking a balance between straggler mitigation and communication costs. Enter the DecantFed algorithm (\ul{D}ynamic cli\ul{e}nt \ul{c}lustering, b\ul{an}dwidth allocation, and local \ul{t}raining for semi-synchronous \ul{Fed}erated learning), a dynamic solution that optimizes client clustering, bandwidth allocation, and local training workloads to maximize data sample processing rates. Additionally, DecantFed adapts client learning rates according to their tiers, addressing the model staleness problem. The algorithm's performance shines in extensive simulations using benchmark datasets, including MNIST and CIFAR-10, under independent and identically distributed (IID) and non-IID scenarios. DecantFed outpaces FedAvg and FedProx in terms of convergence speed and delivers a remarkable minimum 28$\%$ boost in model accuracy compared to FedProx.

\end{abstract}

\IEEEpeerreviewmaketitle

\section{Introduction}
With billions of connected Internet of Things (IoT) devices being deployed in our physical world, IoT data are produced in a large volume and high velocity \cite{Ansari:2018:MEC}. Analyzing these IoT data streams is invaluable to various IoT applications that can provide intelligent services to users \cite{Sun:2016:EdgeIoT}. However, most IoT data streams contain users' personal information, and so users are not willing to share these IoT data streams with third parties, but keep them locally, thus leading to data silos. To break data silos without compromising data privacy, federated learning (FL) has been proposed to enable IoT devices to collaboratively and locally train machine learning models without sharing their data samples \cite{pmlr-v54-mcmahan17a}. 

In general, there are four steps in each global iteration during the FL process. 1) Global model broadcasting, i.e., an FL server initializes a global model and broadcasts it to all the selected clients via a wireless network. 2) Local model training, i.e., each selected client trains the received global model over its local data samples to derive a local model based on, for example, stochastic gradient descent (SGD). 3) Local model update, i.e., each selected client uploads its local model to the FL server via the wireless network. 4) Global model update, the FL server aggregates all the received local models to generate a new global model for the next global iteration. The global iteration continues until the global iteration is converged. 
A typical example of using FL is training a next-word prediction model, where FL enables different smartphones to collaboratively train their next-word prediction models based on messages in their local storage without uploading them to a centralized controller \cite{pmlr-v54-mcmahan17a}. 

The conventional FL, which randomly selects a number of clients to participate in the FL process, may lead to the straggler problem \cite{Vu:2021:SEM}. Here, stragglers refer to the clients with low computation capability that spend a long time for model training or the clients with a low communication data rate that spend a long time for model uploading. So, the FL server has to wait a long time for these stragglers in computing and uploading their local models to the FL server, thus leading to a long training time (which equals the sum of the latency for all the global iterations). 
In order to reduce the waiting time of the FL server, there are two major solutions to resolve the straggler problem, i.e., synchronous FL and asynchronous FL. 
1) In synchronous FL, the FL server sets up a deadline $\tau$, and so it only selects the qualified clients, who can finish its local model training and uploading before the deadline, to participate in the FL process. All the local models that are received after the deadline will be discarded \cite{Rana:2021:APS}. 
Synchronous FL, however, has its own drawbacks. First, synchronous FL reduces the number of participated clients, thus slowing down the convergence rate in terms of requiring more global iterations to converge \cite{wang2018cooperative} even if it can bound the latency of each global iteration to no larger than the deadline. Second, synchronous FL may cause model overfitting owing to data sample diversity reduction. That is, the derived model can fit well for the data samples in these qualified clients, but not in the non-qualified clients, thus reducing the model accuracy. 
2) In asynchronous FL, the FL server enables all the clients to participate in the process and it immediately updates the global model once it receives one from a client. The updated global model would only be sent to that client, who just uploads its local model \cite{8761315}. Asynchronous FL can reduce the waiting time of the FL server, but it would suffer from i) the high communication cost since both the FL server and clients will more frequently exchange their models, and ii) the model stale problem, i.e., slow clients were trained based on an outdated global model, thus reducing the convergence rate \cite{xu2021asynchronous,Damaskinos:2020:Fleet}. 

\begin{figure*}[!htb]
	\centering	
	\includegraphics[width=\textwidth]{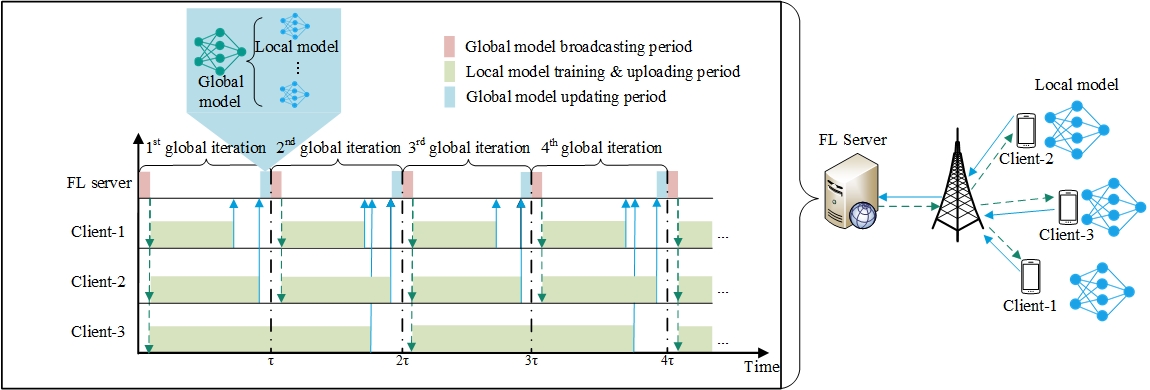} 
	\caption{The semi-synchronous FL framework.}
	\label{fig:sim_FL_overview} 
\end{figure*}

To address the limitations of both synchronous and asynchronous FL, we propose semi-synchronous FL. 
 In this method, all clients in the system are grouped into different tiers based on their model training and uploading latency, and the deadline of a global iteration is denoted as $\tau$. Specifically, if a client can complete model training and uploading within the interval of $\tau\times (j-1)$ and $\tau \times j$, it is assigned to the $j^{th}$ tier.
 As illustrated in Fig. \ref{fig:sim_FL_overview}, we can see an example where Client-1 and Client-2 complete their model training and uploading before the deadline $\tau$, and hence, they are assigned to the $1^{st}$ tier, which has a deadline of $\tau$. On the other hand, Client-3 completes its local model training and uploading between $\tau$ and $2\tau$, which places it in the $2^{nd}$ tier, with a deadline of $2 \tau$. The different tiers imply that clients have different frequencies in uploading their local models to the FL server. For instance, the upload frequency of Client-1 and Client-2 is twice as high as that of Client-3.

Semi-synchronous FL is different from synchronous FL since all the clients in semi-synchronous FL participate in the training process to increase the diversity of training samples, avoid model overfitting and stragglers, and reduce the convergence time. Semi-synchronous FL is also different from asynchronous FL since different clients in semi-synchronous FL have to follow their own schedules to upload their local models and the FL server still has to wait for a global iteration to terminate in order to aggregate the received local models, which reduces the communication cost and mitigate the model stale problem in asynchronous FL.

Although semi-synchronous FL can potentially resolve the drawbacks of synchronous and asynchronous FL, it has its own unveiled challenges. First, how to cluster clients into different tiers? Client clustering depends on the model training and uploading latency of a client. However, it is not easy to estimate the uploading latency of a client, which, for example, depends on the number of clients in this tier. In particular, more clients in the tier sharing the bandwidth resource leads to higher uploading latency and vice versa. Hence, it is hard to estimate the uploading latency if client clustering has not been determined. Second, how to dynamically assign computing workloads to different clients? 
In conventional FL, clients perform uniform local training, meaning that clients train their local models based on the same number of data samples. This results in high-computing-capacity clients having long idle time in each global iteration. For example, as shown in Fig. \ref{fig:sim_FL_overview}, Client-1 in tier-1 has a high computing capacity to quickly finish its local model training, upload its local model to the FL server,  and then wait for the beginning of a new global iteration. Instead of waiting, Client-1 can train more data samples to generate a better local model. 
Studies have demonstrated that training more data samples per client can accelerate model convergence in FL \cite{9771929,Albaseer2022,MLSYS2020_38af8613}. Thus, dynamically optimizing the training workload of a client based on its computing capacity can potentially improve the FL performance. However, workload optimization and client clustering are coupled together since a different workload (i.e., selecting a different number of data samples to train the model) of a client would change its model computing and uploading latency, which leads to a different client clustering. The paper aims to resolve the mentioned two challenges by jointly optimizing client clustering, bandwidth allocation, and workload assignment in the context of semi-synchronous FL.
The main contributions of the paper are summarized as follows.
\begin{enumerate}
    \item We propose semi-synchronous FL to resolve the drawbacks of synchronous and asynchronous FL.   
    \item We propose dynamic workload optimization in semi-synchronous FL and prove that dynamic workload optimization outperforms uniform local training in semi-synchronous FL via extensive simulations.
    \item To resolve the challenges in semi-synchronous FL, we formulate an optimization problem and design 
    \ul{D}ynamic cli\ul{e}nt \ul{c}lustering, b\ul{an}dwidth allocation, and workload op\ul{t}imization for semi-synchronous \ul{Fed}erated learning (DecantFed) algorithm to solve the problem.
\end{enumerate}
The rest of the paper is organized as follows. The related work is summarized in Section II. The system models and the problem formulation are presented in Section III. Section IV describes the proposed DecantFed algorithm to efficiently solve the problem. 
Simulation results are provided and analyzed in Section V, and we conclude the paper in Section VI.

\section{Related Work} 
To solve the straggler problem in FL, synchronous FL has been proposed to set up a deadline and discard the local models received after the deadline. Since having more clients uploading their models can potentially accelerate the training speed, Nishio \emph{et al.} \cite{8761315} proposed to select the maximum number of clients, who can upload their local models before the deadline. On the other hand, since the latency of a client in uploading its model depends on the allocated bandwidth, wireless resource management and client selection are usually optimized jointly. For example, Albaseer \emph{et al.} \cite{Albaseer2022} aimed to jointly optimize the bandwidth allocation and client selection such that the weighted number of the selected clients is maximized. Given a set of selected clients, 
Li \emph{et al.} \cite{MLSYS2020_38af8613} proposed FedProx to assign more local epochs to the clients with high computing capacity, where more local epochs imply higher computing workload. FedProx can potentially reduce the number of global iterations but may hurt the global model convergence owing to local model overfitting, especially when data sample distribution is non-independent and identically distributed (non-IID) \cite{bonawitz2019federated}. FedProx adds a proximal term in the loss function for each client's local model to mitigate the local model overfitting problem. 
Albelaihi \emph{et al.} \cite{Rana:2021:APS} considered time division multiple access (TDMA) as the wireless resource-sharing approach in wireless FL. While TDMA can enhance bandwidth utilization, it can also lead to a waiting period for the selected client to upload their local model due to the wireless channel's limited availability. To account for the potential impact of waiting time on client selection, a heuristic algorithm is developed to maximize the number of selected clients who can upload their models before the deadline.
Selecting more clients to participate in each global iteration leads to the high energy consumption of the clients \cite{10001569}, and so Yu \emph{et al.} \cite{9509751} designed a solution to maximize the tradeoff between the total energy consumption of the selected clients and the number of the selected clients in each global iteration. Instead of maximizing the number of selected clients in each global iteration, Xu and Wang \cite {Xu:2021:CSB} proposed that increasing the number of the selected clients over global iterations would have a faster convergence rate. Based on this hypothesis, they designed a long-term client selection and bandwidth allocation solution, which aims to maximize the weighted total number of the selected clients for all the global iterations, while satisfying the energy budget of the clients and bandwidth capacity limitation. Here, the weight is increasing over the global iterations. 

Instead of updating the global model after the deadline expires in synchronous FL, asynchronous FL has been proposed to enable the FL server to immediately update the global model once it receives a local model \cite{xu2021asynchronous,pmlr-v151-nguyen22b,9725259}. However, as we mentioned before, asynchronous FL has the model stale problem, i.e., the slower clients may train their local models based on an outdated global model, which results in slow convergence or even leads to model divergence \cite{10.1145/3035918.3035933,Wei:2016:SAD}. To mitigate the model stale problem, Xie \emph{et al.} \cite{Xie:2019:AFO} proposed a new global model aggregation, where the new global model equals the weighted sum of the old global model and the received local model. Here, the weight of a local model is a function of the local model staleness, i.e., a longer delay of training and uploading a local model for a client leads to a higher staleness of the local model and a lower weight of the local model. Chen \emph{et al.} \cite{9378161} proposed the asynchronous online FL method, where different clients collaboratively train a global model over continuous data streams locally. The FL server updates the global model once it receives the gradient, learning rate, and the number of data samples from a client. Another drawback of asynchronous FL is the high communication cost, especially for faster clients since they more frequently exchange their models with the FL server. A potential solution could be to enable the FL server to update the global model after it receives a number of local models instead of receiving each local model.    

\section{System Models and Problem Formulation}
In this section, we will provide the related system models and formulate the problem for jointly optimizing client clustering, bandwidth allocation, and workload optimization in semi-synchronous wireless FL. Denote $\bm{\mathcal{I}}$ as the set of clients in the system, and $i$ is used to index these clients. All the clients in $\bm{\mathcal{I}}$ will be clustered into different tiers and will upload their local models to the FL server via a base station (BS). Denote $\bm{\mathcal{J}}$ as the set of tiers, and $j$ is used to index these tiers. 
In our scenario, we consider frequency division multiple access (FDMA) as the wireless resource-sharing mechanism among various tiers. This implies that each tier will be allotted a distinct frequency band that does not overlap with other tiers, and the bandwidth allocated to each tier will be optimized.
Let $b_j$ be the bandwidth allocated to tier $j$. In addition, the clients within the same tier would share the same frequency band based on TDMA, meaning that the clients would take turns acquiring the entire frequency band to upload their models. As mentioned in Section II, TDMA may introduce extra waiting time for a client to wait for wireless bandwidth to be available. Therefore, if client $i$ is in tier $j$, then the latency of client $i$ in training and uploading its local model to the FL server is
\begin{equation}
    t_{ij}=t_i^{comp}+t_{ij}^{wait}+t_{ij}^{upload},
\label{eq:client_latency}
\end{equation}
where $t_i^{comp}$ is the computing latency of client $i$ in training its local model, $t_{ij}^{wait}$ is the delay of client $i$ in tier $j$ to wait for the frequency band to be available, and $t_{ij}^{upload}$ is the uploading latency of client $i$ in tier $j$ to upload its local model.

\subsection{Computing latency}
The computing latency of client $i$ can be estimated by \cite{yang2020delay}
\begin{equation} \small
t^{comp}_i =\frac{C_i  d_i} {f_i}, 
\label{eq:comp_latency}
\end{equation}
where $C_i$ is the number of CPU cycles required for training one data sample of its local model, $d_i$ is the number of samples to be trained on client $i$; and $f_i$ is CPU frequency of client $i$.

\subsection{Uploading latency}
The achievable data rate of client $i$ in tier $j$ can be estimated by
$
{r_{ij}} = b_j\log_2\left( {1 + \frac{{{p_i}g_i}}{{{N_0}}}} \right),
$
where $b_j$ is the amount of bandwidth allocated to clients in tier $j$, ${p_i}$ is the transmission power density of client $i$, ${g_i}$ is the channel gain from client $i$ to the BS, and ${N_0}$ is the average background noise. 
The latency of client $i$ in tier $j$ to upload its local model with size of $s$ to the BS is $t^{upload}_{ij} =\frac{s}{r_{ij}}=\frac{s}{b_{j}\log_2\left(1+\frac{p_ig_i}{N_0}\right)}$.
\subsection{Waiting time}
Once client $i$ finishes its local model training, it can immediately upload its local model to the FL server if the bandwidth is unoccupied, i.e., there is no other client currently uploading its local model. However, if the bandwidth is occupied, client $i$ has to wait, and the waiting time depends on its position in the queue.
Specifically, denote $\bm{\mathcal{I}}_j$ as the set of clients in tier $j$, 
who are sorted based on the increasing order of their computing latency, i.e., $\forall i' \in \bm{\mathcal{I}}_j$, $t_{i'}^{comp}\le t_{i'\!+\!1}^{comp}$, where $i'$ is the index of the clients in $\bm{\mathcal{I}}_j$. Hence, if client $i$ is in tier $j$, its waiting time can be calculated based on the following recursive function. 
\begin{equation}\small
t^{wait}_{i'j}=\begin{cases}
	0,&	i'=1,\\
	\max \left\{ 0,t_{i'\!-\!1}^{comp}+t_{i'\!-\!1,j}^{wait}+t_{i'\!-\!1,j}^{up}-t_{i'}^{comp} \right\} ,& i'>1,\\
\end{cases}
\label{eq:waiting_time}
\end{equation}

\subsection{Problem formulation}
We formulate the problem to jointly optimize client clustering, bandwidth allocation, and workload management in semi-synchronous FL as follows. 
\begin{align}
\bm{P0}:&\mathop {\arg \max }\limits_{\bm{d},\bm{x},\bm{b}} \sum\limits_{i \in \bm{\mathcal{I}}}\left({d_i} \sum\limits_{j \in \bm{\mathcal{J}}} {w_jx_{ij}}\right),\\
\text{s.t.}\ \ \ \ & \forall i \in \bm{\mathcal{I}},\left(t_{i}^{comp}+t_{ij}^{wait}+t_{ij}^{upload}\right) x_{ij}\le j\tau,\label{avaliability}\\
&\forall i \in \bm{\mathcal{I}}, D^{min} \le d_i ,\label{minsampels}\\
&\forall i \in \bm{\mathcal{I}},  \sum\limits_{i \in \bm{\mathcal{J}}}{x_{ij}} = 1 \label{clustering},\\
&\forall i \in \bm{\mathcal{I}}, \forall j \in \bm{\mathcal{J}}, x_{ij} = \left\{0,1\right\}\label{assignment},\\
&  \sum\limits_{j \in \bm{\mathcal{J}}} b_{j} \le B\label{bandwidth}.
\end{align} 
where $x_{ij}$ is a binary variable to indicate if client $i$ is clustered in tier $j$ (i.e., $x_{ij}=1$) or not (i.e., $x_{ij}=0$), $\bm{x} = \left\{ {{x_{ij}}\left| {\forall i \in \bm{\mathcal{I}}}, {\forall j \in \bm{\mathcal{J}}} \right.} \right\}$, $\bm{d} = \left\{ {{d_i}\left| {\forall i \in \bm{\mathcal{I}}} \right.} \right\}$ is the workload allocation in terms of the number of data samples being used to train local models for all the clients, $\bm{b} = \left\{ {{b_j}\left| {\forall j \in \bm{\mathcal{J}}} \right.} \right\}$ implies the amount of bandwidth allocated to different tiers, and $w_j$ indicates the frequency of the clients in tier $j$ uploading their local models. i.e., 
    $w_j=\frac{\left|\bm{\mathcal{J}}\right|-j+1}{\left|\bm{\mathcal{J}}\right|}$.
Here, $\left|\bm{\mathcal{J}}\right|$ indicates the last tier. Hence, a larger $w_j$ means tier $j$ is a lower tier and clients in this tier upload their local models more frequently. Training a model over more data samples leads to higher model accuracy. Based on this intuition, we set up the objective of $\bm{P0}$ to maximize the number of training data samples per unit time, where $\sum\limits_{i \in \bm{\mathcal{J}}} {w_jx_{ij}}$ is the frequency of client $i$ in uploading its local model, and $d_i$ is the number of data samples used to train its local model. Constraint \eqref{avaliability} indicates the clients in different tiers should meet their tier deadlines. Constraint \eqref{minsampels} defines the minimum number of the samples, denoted as $D^{min}$, to be trained locally by each client. Constraint \eqref{clustering} specifies each client should only be clustered into exactly one tier. Constraint \eqref{assignment} refers to $x_{ij}$ being a binary variable, and Constraint \eqref{bandwidth} means the amount of bandwidth allocated to all the tiers no larger than $B$, where $B$ is the available bandwidth at the BS.      



\section{The DecantFed algorithm}
$\bm{P0}$ is difficult to be solved. First, client clustering $\bm{x}$, bandwidth allocation $\bm{b}$, and workload management $\bm{d}$ are coupled together. For instance, allocating less bandwidth or selecting more data samples for client $i$ in a tier would increase its uploading latency $t^{upload}_i$ and computing latency $t^{comp}_i$, thus leading to client $i$ unable to upload its local model before its tier deadline. As a result, client $i$ has to be assigned to the next tier, which has a longer tier deadline. 
We propose DecantFed to efficiently solve problem $\bm{P0}$. The basic idea of DecantFed is to decompose $\bm{P0}$ into two sub-problems and solve them individually. 

\subsection{Client clustering and bandwidth allocation}
Assume that each client trains only the minimum number of samples $D^{min}_i$ from its local dataset, i.e., for all $i$ in $\bm{\mathcal{I}}$, $d_i=D^{min}$, then $\bm{P0}$ can be transformed into the following.

\begin{align}
\bm{P1}:&\mathop {\arg \max }\limits_{\bm{x},\bm{b}} \sum\limits_{i \in \bm{\mathcal{I}}} \sum\limits_{j \in \bm{\mathcal{J}}} {w_jx_{ij}},\\
\text{s.t.}\ \ \ \ & \text{Constraints}\ \eqref{avaliability},\ \eqref{clustering},\ \eqref{assignment},\ \text{and}\  \eqref{bandwidth} \nonumber.
\end{align} 
Solving $\bm{P1}$ remains challenging due to the unknown waiting time $t^{wait}_{ij}$ specified in \eqref{avaliability}, which becomes known only after clients have been grouped into tiers. However, for optimizing the objective function in $\bm{P1}$, it is preferable to assign clients to lower tiers that have a larger $w_j$, while still meeting the tier's deadline constraints as outlined Constraint \eqref{avaliability}. 
Motivated by this insight for solving $\bm{P1}$, we design the heuristic LEAD algorithm (joint cLient clustEring and bAnDwidth allocation), summarized in Algorithm \ref{alg:P1}. Specifically,
\begin{enumerate}
    \item{Each client's uploading time is determined by its tier. Additionally, tier bandwidth is determined by the number of clients in the corresponding tier, i.e., 
    $b_j=\frac{\sum\limits_{i \in \bm{\mathcal{I}}} {x_{ij}}}{\left|\bm{\mathcal{I}}\right|}\times B$. Then, 
     \begin{equation} \small
    t^{upload}_{ij} =\frac{\left|\bm{\mathcal{I}}\right|}{B\sum\limits_{i \in \bm{\mathcal{I}}} {x_{ij}}}\frac{s}{\log_2\left(1+\frac{p_ig_i}{N_0}\right)}.
    \label{eq:upload_latency_tier}
    \end{equation}
  }
    
    \item{In the $j^{th}$ iteration, we intend to assign as many clients as possible to the $j^{th}$ tier, while satisfying Constraint \eqref{avaliability}. Denote $\tilde{\bm{\mathcal{I}}}$ as the set of clients, who have not been clustered into any tier yet, i.e., $\tilde{\bm{\mathcal{I}}}=\left\{ \forall i\,\,\in \bm{\mathcal{I}}\left| \sum_{j\in \bm{\mathcal{J}}}{x_{ij}}=0 \right. \right\}$. We sort the clients in $\tilde{\bm{\mathcal{I}}}$ in descending order based on their computing latency. Denote $\bm{\mathcal{I}}_j$ as the set of these sorted clients, i.e., $\bm{\mathcal{I}}_j=\left\{ \forall i' \in \tilde{\bm{\mathcal{I}}}\left| t_{i'+1}^{comp}\le t_{i'}^{comp} \right. \right\}$, where $i'$ is the index of the clients in $\bm{\mathcal{I}}_j$. Assume that all the clients in $\bm{\mathcal{I}}_j$ can be assigned in tier $j$, i.e., $\forall i' \in {\bm{\mathcal{I}}}_j, x_{i'j} = 1$, and then 
    we iteratively check the feasibility of all the clients in tier $j$ starting from the first client, i.e., whether each client in tier $j$ can really be assigned to tier $j$ to meet Constraint \eqref{avaliability} or not. If a client currently in tier $j$ cannot meet Constraint \eqref{avaliability}, this client will be removed from tier $j$, i.e., $x_{i'j}=0$. Note that removing a client reduces the bandwidth $b_j$ allocated to tier $j$ in Step 13, which may lead to the clients, who were previously feasible to be assigned to tier $j$ to meet Constraint \eqref{avaliability}, no longer feasible because of the decreasing of $b_j$. As a result, we have to go back and check the feasibility of all the clients in tier $j$ starting from the first client again, i.e., $i'=1$ in Step 14. 
    }
    \item{Once the client clustering in tier $j$ is finished, we start to assign clients to tier $j+1$ by following the same procedure in Steps 5-19. The client clustering ends when all clients have been assigned to the existing tiers, i.e., $\tilde{\bm{\mathcal{I}}} \ne \emptyset$.}
\end{enumerate}\

\begin{figure}[!t]
\removelatexerror
\begin{algorithm}[H]
\label{alg:P1}
\SetAlgoLined
\caption{LEAD algorithm}
    $\forall i \in \bm{\mathcal{I}}, d_i=D^{min}_i$, calculate $t^{comp}_i$ based on Eq.~\eqref{eq:comp_latency}.
    

   $\forall i \in \bm{\mathcal{I}}, \forall j \in \bm{\mathcal{J}}$, initialize $x_{ij} = 0$.
   
   Initialize $j=1$ and $\tilde{\bm{\mathcal{I}}}:=\bm{\mathcal{I}}$.
   
   
   \While {$\tilde{\bm{\mathcal{I}}} \ne \emptyset$}
   {
   
   Create ${\bm{\mathcal{I}}}_j$ by sorting the clients in $\tilde{\bm{\mathcal{I}}}$ based on their computing latency; 

   Cluster all the clients in ${\bm{\mathcal{I}}}_j$ into tier $j$, i.e., $\forall i' \in {\bm{\mathcal{I}}}_j, x_{i'j} = 1$;
   
   
   Calculate $b_j=\frac{\left|\bm{\mathcal{I}}_j\right|}{\left|\bm{\mathcal{I}}\right|}\times B$;
   
   
   
    \For {$i'\xleftarrow{}1$ to ${\left|\bm{\mathcal{I}}_j\right|}$}
   {
    \If{$x_{i'j}==1$}{
   
   Calculate $t_{i'j}^{wait}$ and $t_{i'j}^{upload}$ based on Eqs. \eqref{eq:waiting_time} and \eqref{eq:upload_latency_tier}, respectively;
   
   \If{$t_{i'}^{wait}\!+\!t_{i'j}^{wait}\!+\!t_{i'j}^{upload}\!>\!\tau j$} {
   
   $x_{i'j} = 0$;
   
   
   Update $b_j=\frac{\sum\limits_{i' \in \bm{\mathcal{I}}_j} {x_{i'j}}}{\left|\bm{\mathcal{I}}\right|}\times B$;
   
   Restart the for loop, i.e., $i'=1$.
   }
   }
   }

   $\tilde{\bm{\mathcal{I}}}=\left\{ \forall i\,\,\in \bm{\mathcal{I}}\left| \sum_{j\in \bm{\mathcal{J}}}{x_{ij}}=0 \right. \right\}$;
   
   $j:=j+1;$

   }
\end{algorithm}
\end{figure}

\begin{figure}[!t]
 \removelatexerror
\begin{algorithm}[H]
\label{alg:Decant}
\SetAlgoLined
\caption{DecantFed algorithm}
    
    \SetKwBlock{Fnb}{\textnormal{\textbf{At the FL server}:}}{end}{
    
    \Fnb{
    
 \SetKwBlock{kk}{\textnormal{\textbf{Client/Task scheduling (performed once)}:}}{end}{
    \kk{
    Obtain states of all clients, i.e., $\forall i \in {\bm{\mathcal{I}}}$, $g_i$, $f_i$, $p_i$
    
    Derive $\bm{b}$ and $\bm{x}$ by executing the LEAD algorithm. 
    
    Derive $\bm{d}$ by applying the Simplex to solve $\bm{P2}$. 

    Broadcast the workload assignment $\bm{d}$, bandwidth allocation $\bm{b}$, client clustering $\bm{x}$, deadline $\tau$, and the initial global model to all the clients.

}}
    \For{incremental global iteration $l$}{
    Receive local models from the clients in ${\bm{\mathcal{K}}_l}$;
    
    Update the global model ${\bm{\omega}}_l$ based on Eq. \eqref{eq:global};
    
    Broadcast the updated global model ${\bm{\omega}}_{l}$ to the clients in ${\bm{\mathcal{K}}_l}$; }
    }
    }

\SetKwBlock{Fnb}{\textnormal{\textbf{At each client}:}}{end}{
    
    \Fnb{Receive $\bm{d}$, $\bm{b}$, $\bm{x}$, $\tau$, and the initial global model to understand the local training configuration.

    Calculate the learning rate based on Eq. \eqref{eq:learning_rate}.
    
    \While{a new global model is received}{ 
    Train its local model over $d_i$ samples based on Eqs. \eqref{eq:gradient_decent} and \eqref{eq:clipping};
    
    Upload its local model to the FL server;

    }
    }
    }

\end{algorithm}
\end{figure}
%
%
%
 %
%
%
 %
%
%
%
%
%
%
%
%
%
%
\subsection{Dynamic workload optimization}
Plugging client clustering $\bm{x}$ and bandwidth allocation $\bm{b}$ that are derived by the LEAD algorithm into $\bm{P0}$, we have 
\begin{align}
\bm{P2}:
&\mathop {\arg \max }\limits_{\bm{d}} \sum_{i\in \bm{\mathcal{I}}}{\left( \sum_{j\in \bm{\mathcal{J}}}{w_jx_{ij}} \right)}d_i,\\
\text{s.t.}\ \ \ \ & \text{Constraints}\ \eqref{avaliability}\ \text{and}\  \eqref{minsampels} \nonumber.
\end{align} 
Assume that $\forall i \in \bm{\mathcal{I}}$, $d_i$ is a continuous variable. Then, $\bm{P2}$ is a linear programming problem, and its optimal solution can be derived by using the Simplex method in polynomial time \cite{dantzig_thapa_2010}. Note that the number of samples $d_i$ should be an integer variable, and so we simply take $d_i$ to be the maximum integer value that is smaller than $d^*_i$, i.e.,  $\forall i \in \bm{\mathcal{I}}$, $d_i=\lfloor d_{i}^{*} \rfloor$, where $d^*_i$ is the optimal value by using the Simplex method to solve $\bm{P2}$.

\subsection{Summary of DecantFed}
In DecantFed, the FL server 1) obtains the states in terms of channel gain $g_i$, CPU frequency $f_i$, and transmission power $p_i$ of all the clients, 2) calculates the bandwidth allocation $\bm{b}$ and client clustering $\bm{x}$ based on LEAD in Algorithm \ref{alg:P1} as well as the workload $\bm{d}$ by using Simplex method to solve $\bm{P2}$, 3) broadcasts the workload assignment $\bm{d}$, bandwidth allocation $\bm{b}$, client clustering $\bm{x}$, deadline $\tau$, and the initial global model to all the clients such that all the clients can understand their local training configurations, and 4) conducts FL training for a number of global iterations. Here, in the $l^{th}$ global iteration, the FL server only collects models from the clients, whose tiers are participating in training during the $l^{th}$ global iteration. That is, if $l\%j==0$, then tier $j$ is participating in training during the $l^{th}$ global iteration. Denote ${\bm{\mathcal{K}}_l}$ as the set of the clients that are participating in the training during the $l^{th}$ global iteration, i.e., ${\bm{\mathcal{K}}_l}=\{i|\forall i \in {\bm{\mathcal{I}}}, \forall j \in {\bm{\mathcal{J}}}, l\%j==0\ \&\&\ x_{ij}==1 \}$. Once the FL server receives the local models from all the clients in ${\bm{\mathcal{K}}_l}$, it then updates the global model based on 
\begin{equation}
\label{eq:global}
\bm{\omega }_{l}^{global}=\sum_{i\in \bm{\mathcal{K}}_l}{\frac{\left| \bm{\mathcal{D}}_i \right|}{\sum_{k\in \bm{\mathcal{K}}_l}{\left| \bm{\mathcal{D}}_k\right|}}}\bm{\omega}_{i,l}^{local},
\end{equation}
where $\bm{\omega}_{i,l}^{local}$ is the local model uploaded from client $i$ in the $l^{th}$ global iteration, $\bm{\omega }_{l}^{global}$ is the global model derived in the $l^{th}$ global iteration, ${\bm{\mathcal{D}}_{i}}$ is the local data set of client $i$, and $\left| \bm{\mathcal{D}}_i \right|$ is the total number of samples in ${\bm{\mathcal{D}}_{i}}$. Once the FL server has updated the global model $\bm{\omega }_{l}^{global}$ based on \eqref{eq:global}, it broadcasts $\bm{\omega }_{l}^{global}$ to the clients in ${\bm{\mathcal{K}}_l}$ and then initializes the next global iteration. 


In DecantFed, each client trains the received global model based on its $d_i$ number of local samples. However, the local training may have 1) the model staleness problem due to the semi-synchronous nature, and 2) the model divergence problem due to the non-IID and workload optimization. Here, we apply the following solutions to tackle the model staleness and divergence problems.
\begin{enumerate}
\item {\textbf{Dynamic learning rate}. Similar to asynchronous FL, the model staleness problem may also exist in DecantFed, although it is mitigated. The reason for having staleness in DecantFed is that clients in high tiers may train their local models based on outdated global models. To further mitigate the model staleness problem in DecantFed, we adopt the method in \cite{9378161} to set up different learning rates for the clients in different tiers, i.e.,
\begin{equation}
\delta_j=\min\left(\delta_{1}\times \max(\log_\alpha j,1),0.1\right)
\label{eq:learning_rate}
\end{equation}
where $\delta_j$ is the learning rate of the clients in tier $j$,  $\delta_1$ is the learning rate of the clients in tier 1, and $\alpha$ ($\alpha > 1$) is a hyperparameter to adjust the changes of learning rates among tiers. Note that both $\delta_1$ and $\alpha$ are given before the FL training starts. Basically, Eq. \eqref{eq:learning_rate} indicates that clients in a higher tier would have a larger learning rate (i.e., $\delta_{j+1} \ge \delta_j$) to update their local models more aggressively such that these clients can catch up with the model update speed of the clients in the lower tiers. Note that Eq. \eqref{eq:learning_rate} also ensures that $\delta_j$ is less than 0.1 to avoid the clients in higher tiers overshooting the optimal model that minimizes the loss function.
%
Once the learning rate is determined, client $i$ updates its local model based on the gradient descent method, i.e.,
\begin{align}
\bm{\omega }_{i,l}^{local}\left( t+1 \right)&=\bm{\omega }_{i,l}^{local}\left( t \right)-\sum_{j\in {\bm{\mathcal{J}}}}{\delta _jx_{ij}} \nonumber\\
&\times \nabla f\left( \bm{\omega }_{i,l}^{local}\left( t \right);a_{i,n},b_{i,n} \right),
\label{eq:gradient_decent}
\end{align}
where $\bm{\omega }_{i,l}^{local}\left( t \right)$ is the local model of client $j$ during the $t$ and $t-1$ local iterations, respectively, in the $l^{th}$ global iteration, $\left(a_{i,n},b_{i,n} \right)$ is the input-output pair for client $i$'s $n^{th}$ data sample, $\sum_{j\in {\bm{\mathcal{J}}}}{\delta _jx_{ij}}$ implies the learning rate of client $i$, and $f\left( \bm{\omega }_{i,l}^{local}\left( t \right); a_{i,n},b_{i,n} \right)$ is the loss function of the model given the model parameter $\bm{\omega }_{i,l}^{local}\left( t \right)$ and data sample $\left(a_{i,n},b_{i,n} \right)$. A typical loss function that has been widely applied in image classification is cross-entropy loss.} 
\item {\textbf{Clipping the loss function values}. Owing to the non-IID and dynamic workload optimization, the data samples in a client could be highly uneven.
For example, a client has 1,000 images labeled as a dog, but only 2 images labeled as a cat. If this client has high computing capability and the FL server would assign a high workload in terms of training a machine-learning model to classify dogs and cats by selecting all the images for many epochs, then the local model may overfit the client's local data set. After being trained by numerous dog images, the local model may diverge if a cat image is fed into the local model to generate an excessive loss value. For example, if the loss function is defined as the cross-entropy loss, then the loss function is $-\log(\epsilon)\!\rightarrow\! \infty$, where $\epsilon\!\rightarrow\!0$ is the probability of labeling the image as a cat by the local model. 
Infinite loss values lead to large backpropagation gradients, which can subsequently turn both weights and biases into `NaN'. 
Although regularization methods can reduce the variance of model updates, especially in IID scenarios, they cannot resolve the infinite loss issues that normally happen when the data distribution is highly non-IID. In such cases, loss clipping serves as an effective and computationally efficient solution to constrain model update, clipping the loss value into a reasonable range, i.e., 
\begin{equation}
f\left( \bm{\omega }_{i,l}^{local}\left( t \right) ;a_{i,n},b_{i,n} \right) :=\min \left\{ f\left( \bm{\omega }_{i,l}^{local}\left( t \right) ;a_{i,n},b_{i,n} \right) , \zeta \right\}, 
\label{eq:clipping}
\end{equation}
where $\zeta$ is a hyperparameter that defines the maximum loss function value. 
}
\end{enumerate}

\section{Simulations}

\subsection{Simulation setup}

\begin{table}[!htb]
	\centering
	\caption{Simulation parameters}
	\label{table_1}
	\begin{tabular}{ll}
		\toprule
		\textbf{Parameter} & \textbf{Value} \\
		\midrule
	    Background noise $N_0$ & $-94$ dBm\\
     Bandwidth ($b$) & 1 MHz\\
     Client transmission power ($p$)& 0.1 watt\\
	    Size of the local model ($s$)& $100$ kbit\\
     Client $i$ CPU frequency &  $f_i \in U (0.1,1)\times10^9$ Hz\\
\makecell[l]{Number of CPU cycles required for \\training
one sample on client $i$}    & $C_i \in U(1,5)\times10^7$\\
	   Number of local samples ${\left| {\bm{\mathcal{D}}}_i \right|}$  &  Various, Dirichlet distribution\\
	   Number of local epochs  & Various, dynamic local training\\
	   Number of local batch size  & $10$\\
		\bottomrule
	\end{tabular}%
	\label{tab:sim_para}%
\end{table}%

 Assume that there are 100 clients that are randomly distributed over a 2 km $\!\!\times\!\!$ 2 km area to participate in the FL training via a wireless network, where a BS is located at the center of the area to forward global/local models between the FL server and clients. In addition, the path loss between a client and the BS is estimated based on $\phi = 128.1+37.6d_i$, where $\phi$ and $d_i$ are the path loss and distance between the BS and client $i$. If fast fading is not considered, the channel gain $g_i$ between the BS and client $i$ is mainly determined by the path loss, i.e., $g_i=10^{-\frac{\phi_i}{10}}$. Moreover, the transmission power of each client is $0.1$ watt, i.e., $\forall i\in \bm{\mathcal{I}}, p_i=0.1$. The amount of bandwidth available for the BS is $B=1$ MHz. Also, the CPU frequency of a client $f_i$ and the number of CPU cycles to process a single sample $C_i$ are randomly selected, following based on two uniform distributions, i.e., $f_i \in U (0.1,1)\times10^9$ Hz and $C_i \in U(1,5)\times10^7$ CPU cycles. Other simulation parameters are listed in Table \ref{tab:sim_para}.
 
\subsubsection{Non-IID Dataset}
Two benchmark datasets, i.e., CIFAR-10 and MNIST, are used to evaluate the performance of DecantFed. Here, CIFAR-10 is a dataset for image classification with 10 image labels and 6000 images per label. It has 50,000 training images and 10,000 test images. MNIST is a dataset of 28$\times$28 grayscale images of handwritten digits 0-9 with 60,000 training examples and 10,000 test examples. 

The benchmark dataset will be dispatched to 100 clients based on non-IID across different categories, and the Dirichlet distribution is a common choice for generating non-IID data. The probability density function of the Dirichlet distribution is \cite{https://doi.org/10.48550/arxiv.2210.10311}
\begin{equation}
    f(\eta_1,\dots,\eta_{\left | \bm{\mathcal{I}} \right |};\beta)=\frac{\Gamma(\beta \times \left | \bm{\mathcal{I}} \right |)}{\Gamma(\beta)^{\left | \bm{\mathcal{I}} \right |}}\prod_{i=1}^{\left | \bm{\mathcal{I}} \right |}\eta_i^{\beta-1},
\label{eq:Dirichlet}
\end{equation}
where $\left | \bm{\mathcal{I}} \right |$ is the total number of clients in the network, $\eta_i$ is the probability of dispatching an image in a dataset to client $i$, where $\sum_{i=1}^{\left | \bm{\mathcal{I}} \right |}\eta_i=1$, and $\beta$ is a parameter to adjust the degree of non-IID, i.e., a smaller $\beta$ leads to a more non-IID data distribution among the clients. $\Gamma(\beta)$ in Eq. \eqref{eq:Dirichlet} is the Gamma function, i.e.,  $\Gamma(x) = \int_0^\infty t^{x-1}e^{-t}dt$ is Gamma function.

\subsubsection{Global model design}

We will train two different convolutional neural networks (CNNs) for the two benchmark datasets. For CIFAR-10, we apply a 3 VGG-block CNN with 32, 64, and 128 filters in the convolution layers. Followed by convolution layers, two fully connected layers with 128 and 10 nodes, respectively, are added, and ReLU is used as the activation function for all these layers. For MNIST, we apply a neural network having two fully connected layers with 784 and 10 nodes, respectively. The learning rates of the clients are calculated based on Eq. \eqref{eq:learning_rate}, where $\alpha=1.45$ and $\delta_1=0.005$. The loss function value is clipped to $\zeta=\log_2(10) \approx 3.33$.  

\subsubsection{Comparison methods} 
Some FL algorithms assume that the FL server understands some prior knowledge, such as the number of samples residing on each client or the distribution of samples in each client, and design corresponding client selection or model aggregation methods to improve the model accuracy or accelerate the training process. However, we argue that this prior knowledge may not be available or accurate and violates data privacy, which is one of the major motivations for applying FL. The proposed DecantFed does not require such prior knowledge, and in order to achieve fair comparisons, two FL baselines, i.e., FedAvg \cite{pmlr-v54-mcmahan17a} and FedProx \cite{MLSYS2020_38af8613}, that also do not require such prior knowledge are used to compare the performance with DecantFed.         
Here, FedAvg selects all the clients in each global iteration without setting up a deadline. Also, FedAvg performs uniform local model training, meaning that the number of data samples to train a local model is the same for all the clients.
 On the other hand, FedProx performs synchronous FL by setting up a deadline $\tau$ and only selects a few clients, who can upload their local models before the deadline, to participate in the model training. Also, FedProx will dynamically allocate workloads to different clients based on their computing capacities. Three FL algorithms are summarized in Table 
 \ref{tab:methods}. To compare the performance of different FL algorithms, an evaluation will be conducted on their convergence rate and model test accuracy. Since the duration of a global iteration may vary for the three FL algorithms, the convergence rates will be measured over both time and global iterations.


\begin{figure*}[!htb]
	\centering	
	\includegraphics[width=\textwidth]{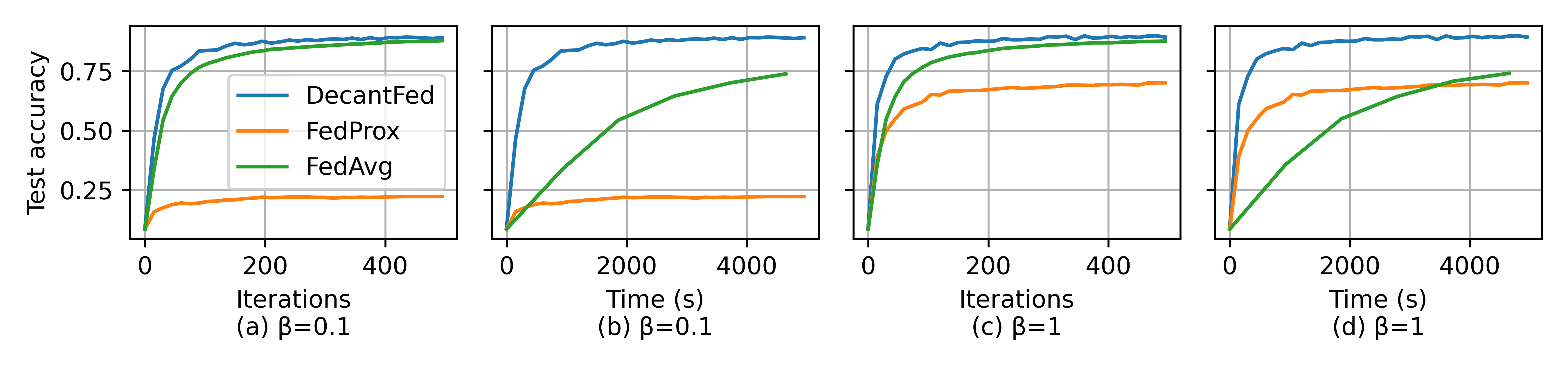} 
	\caption{Test accuracy of different algorithms for MNIST with $\beta=0.1$ and $1$.
	}
	\label{fig:MNIST} 
\end{figure*}

\begin{figure*}[!htb]
	\centering	
	\includegraphics[width=\textwidth]{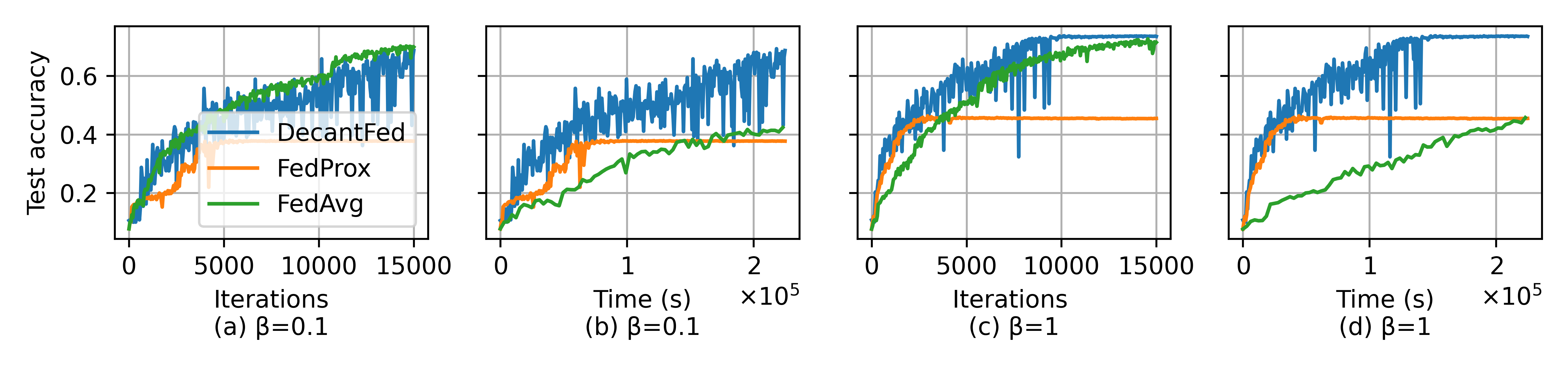} 
	\caption{Test accuracy of different algorithms for CIFAR-10 with $\beta=0.1$ and $1$.
	}
	\label{fig:cifar} 
\end{figure*}


\begin{figure*}[!h]
	\centering	
	\includegraphics[width=\textwidth]{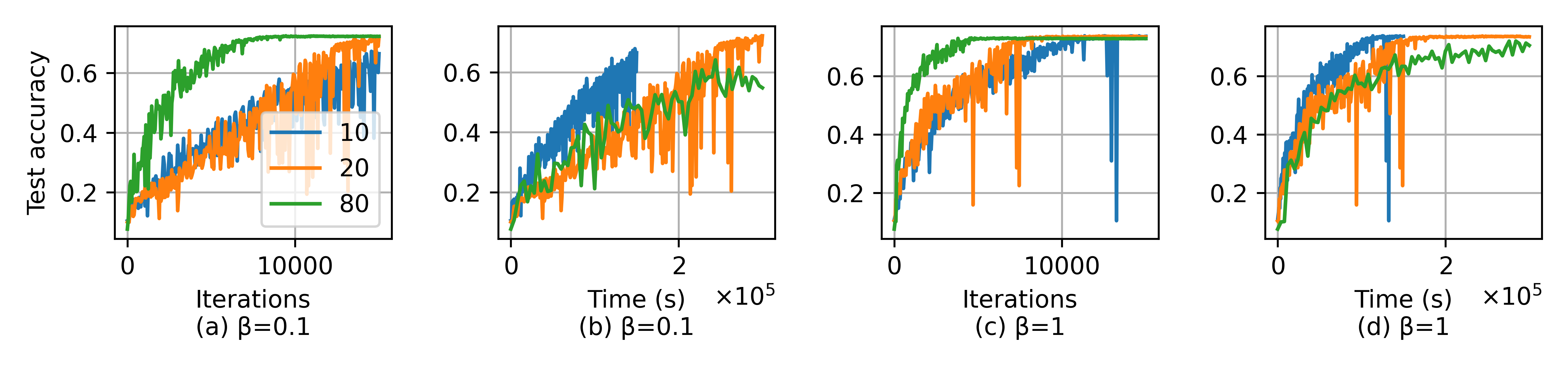} 
	\caption{Test accuracy of  DecantFed with different deadlines for CIFAR-10 with $\beta=0.1$ and $1$.
	}
	\label{fig:deadline} 
\end{figure*}

\begin{figure}[!htb]
	\centering	
	\includegraphics[width=\columnwidth]{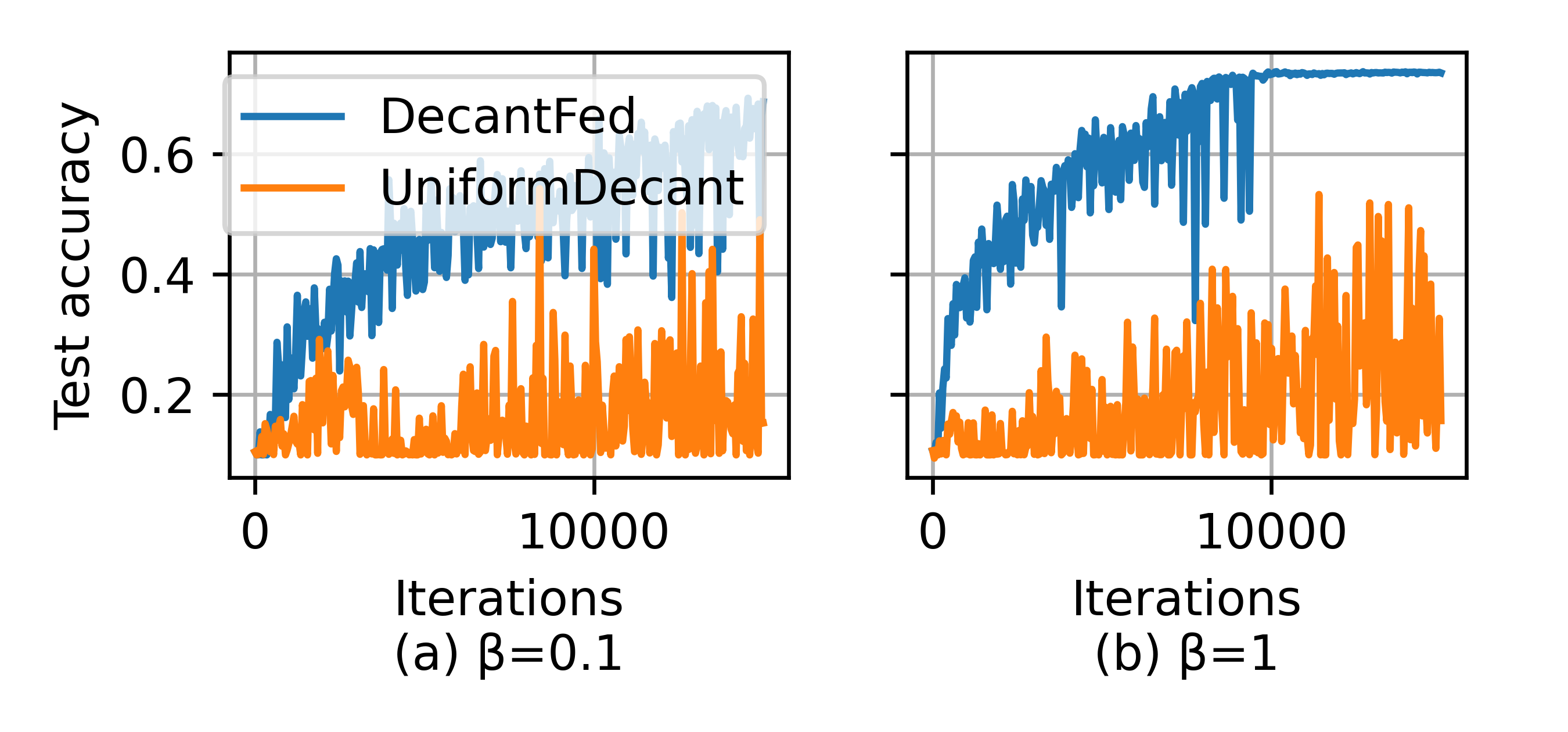} 
	\caption{Dynamical local training}
	\label{fig:dynamic_local_training} 
\end{figure}


\subsection{Simulation results}
\subsubsection{Performance comparison among different FL algorithms}
Assume that the deadline $\tau=15$ seconds. Fig.~\ref{fig:MNIST} shows the learning curves for different algorithms by using MNIST. Here, Figs.~\ref{fig:MNIST}(a) and (b) show the learning curves over global iterations and time, respectively, when the data distribution is more similar to non-IID (i.e., $\beta=0.1$). Similarly, Figs.~\ref{fig:MNIST}(c) and (d) show the learning curves over global iterations and time, respectively, when the data distribution is more similar to IID (i.e., $\beta=1$). From Figs.~\ref{fig:MNIST}(a) and (c), we can see that, in both IID ($\beta=1$) and non-IID ($\beta=0.1$) scenarios, the final model accuracy of DecantFed is slightly higher than FedAvg (which is around $90\%$) but much higher than FedProx, and the convergence rate over global iteration for DecantFed is slightly faster than FedAvg. The reason for having low testing accuracy for FedProx is that FedProx only selects 5 clients out of 100, and low client participation leads to an insufficient and biased training data set, which reduces the testing accuracy of the global model. The low client participation in FedProx also results in model accuracy decreasing when the data distribution changes from IID to non-IID. Yet, the change of model accuracy over $\beta$ for DecantFed and FedAvg is negligible because they allow all the clients to participate in the training.  In Figs.~\ref{fig:MNIST}(b) and (d), it is easy to observe that the convergence rate over time for DecantFed is much faster than FedAvg in both IID and non-IID scenarios. DecantFed is fully converged around $130,000$ seconds achieving $90\%$ model accuracy, while FedAvg only achieves $45\%$ and $73\%$ model accuracy in IID and non-IID scenarios, respectively, at $200,000$ seconds. This is because the duration of a global iteration for FedAvg is much longer than DecantFed since FedAvg has to wait for all the clients to upload the local models.    

Fig.~\ref{fig:cifar} shows the learning curves for different algorithms by using CIFAR-10. We can obtain the similar conclusion that DecantFed and FedAvg converge to a similar model test accuracy, i.e., $70\%$ for non-IID as shown in Fig.~\ref{fig:cifar}(a) and $73\%$ for IID as shown in Fig.~\ref{fig:cifar}(c). On the other hand, FedProx only achieves $39\%$ and $45\%$ in non-IID and IID, respectively. This highlights that DecantFed can achieve at least $28\%$ higher model accuracy than FedProx.
Also, the convergence rate over time for DecantFed is much faster than FedAvg in both IID and non-IID scenarios as shown in Figs.~\ref{fig:cifar}(b) and (d). 
In order to attain a test accuracy of $41\%$, DecantFed and FedAvg demand $0.5\times10^5$ seconds and $2\times10^5$ seconds, respectively. DecantFed is four times faster than FedAvg.
It is worth noting that when $\beta=0.1$ as shown in Fig.~\ref{fig:cifar}(a), the final model accuracy of FedAvg is slightly better than DecantFed, which is different from the other settings. Also, the model accuracy of FedAvg during the training is more stable than DecantFed.         

\subsubsection{Performance of DecantFed by varying $\tau$}

\begin{table}[!htb]
\begin{tabular}{|c|c|c|c|c|}
\hline
Methods & Workload ($d_i$) & Synchronous & Deadline & Clients\\ \hline
DecantFed & dynamic  & semi-syn &  $\tau$& all\\ \hline
FedProx \cite{MLSYS2020_38af8613} & dynamic  & syn &  $\tau$ & few\\ \hline
FedAvg \cite{pmlr-v54-mcmahan17a} & fixed  & syn &  $\infty$& all\\ \hline
\end{tabular}
\caption{Comparisons of different FL algorithms.}
\label{tab:methods}
\end{table}

The deadline $\tau$ plays a critical role in determining the performance of DecantFed. A smaller value of $\tau$ enables more tiers in the system, resulting in fewer clients assigned to a tier. Hence, a smaller $\tau$ makes DecantFed behave more like asynchronous FL, eventually becoming asynchronous when each tier only has one client. Conversely, a larger $\tau$ results in DecantFed behaving more like FedAvg, i.e., all the clients are assigned to a tier and the FL server has to wait for the last client to upload its local model if $\tau \rightarrow \infty$. Fig.~\ref{fig:deadline} shows the learning curve for different deadline settings by using CIFAR-10 when $\beta=0.1$ and $\beta=1$. As shown in Fig.~\ref{fig:deadline} (a) and (c), a larger $\tau$ not only accelerates convergence over iterations but also stabilizes the training process. This is because a larger $\tau$ allows for more low-index-tier clients to contribute to each global update. Additionally, a larger $\tau$ helps global model stability. For example, there are several noticeable sudden drops in test accuracy when $\tau=10$, which is owing to the fact that there is only an average of 1 client in the first tier, likely providing a biased local model update. 
Conversely, when $\tau=80$, the training curve is more stable due to the fact that more clients are in the lower tiers. However, a larger $\tau$ leads to slower convergence over time owing to the longer duration of a global iteration. Table \ref{tab:tau} shows final model accuracy (i.e., when the model is converged) by applying different values of $\tau$ when $\beta=0.1$. We can observe that choosing the appropriate deadline (e.g., $\tau=10$ seconds) is critical to optimize the tradeoff by maximizing the final model accuracy and maximizing the convergence rate over time in DecantFed.     

\begin{table}[!htb]
\begin{tabular}{|c|c|c|c|c|c|c|c|}
\hline
Deadline $\tau$ (s) & 2.5 & 5 & 10 & 20 & 40 &80  \\ \hline
Test accuracy (\%)& 69.07 & 72.59 & 73.48 &  73.45& 73.03& 72.61\\ \hline
\end{tabular}
\caption{Test accuracy for $\tau$s, non-IID $\beta=0.1$.}
\label{tab:tau}
\end{table}


\subsubsection{Performance of DecantFed by optimizing the workload among clients} We next evaluate how the workload optimization affects the performance of DecantFed. Here, we have two settings of DecantFed, where 1) UniformDecant \cite{https://doi.org/10.48550/arxiv.2210.10311} performs uniform local training among the clients, i.e., all the clients use the same number of data samples (i.e., $d_i=10, \forall i \in \bm{\mathcal{I}}$) to train their local models, and 2) DecantFed optimizes $d_i$ by solving $\bm{P2}$. Other settings for UniformDecant and DecantFed are the same. Fig. \ref{fig:dynamic_local_training} shows the learning curves for UniformDecant and DecantFed by using CIFAR-10 when $\beta=0.1$ and $\beta=1$. From the figures, we can find that DecantFed can achieve not only a higher final model accuracy but also a faster convergence rate. 
This basically demonstrates that dynamic workload optimization enables clients with high computational capacities to train their local models over more data samples can significantly improve the FL performance.    



\section{Conclusion}
To solve the drawbacks of synchronous FL and asynchronous FL, we propose a semi-synchronous FL, i.e., DecantFed, which 1) jointly clusters clients into different tiers and allocates bandwidth to different tiers, and so the clients in different tiers would have different deadlines/frequencies to upload their local models; 2) dynamically allocates training workload in terms of training data samples to different clients to enable high computational capacity clients to derive better local models, while keeping the clients in their original tiers. Simulation results demonstrate that the model accuracy incurred by DecantFed and FedAvg is similar, but is much higher than FedProx. The convergence rate over time incurred by DecantFed is much faster than FedAvg. In addition, the deadline is an important parameter that significantly influences the performance of DecantFed, and we will investigate how to dynamically adjust the deadline to maximize the performance of DecantFed in the future. 
Finally, our results demonstrate that dynamic workload optimization for clients is vital to improving FL performance.

\bibliographystyle{IEEEtran}
\bibliography{IEEEabrv,mybibfile}

\begin{IEEEbiography}[{\includegraphics[width=1in,height=1.25in,clip,keepaspectratio]{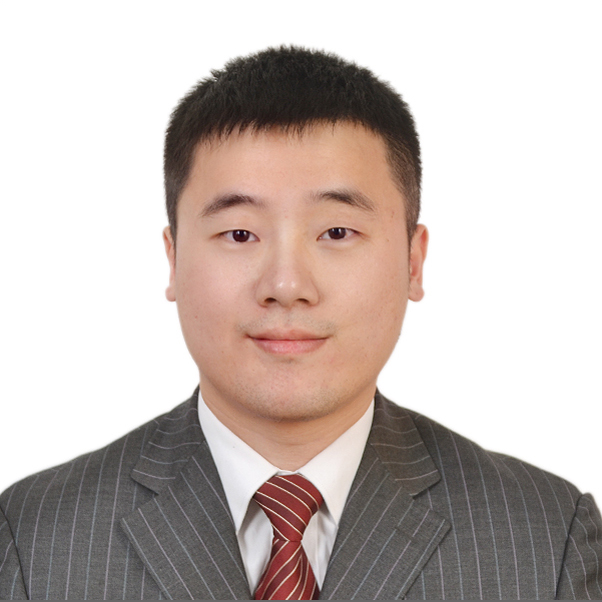}}]{Liangkun Yu}[S'20] obtained his B.E. and M.E. degrees in Communications Engineering from Fuzhou University in 2014 and 2017, respectively. Following this, he joined China Telecom as a wireless network engineer from 2017 to 2019. He commenced his doctoral studies at the SENet lab in the University of New Mexico in 2019. Specializing in machine learning and wireless networks, his research explores various topics, including reinforcement learning for UAV swarm traffic management in aerial corridors, wireless IoT federated learning, queueing theory application in wireless networks, and drone-assisted mobile access networks.
\end{IEEEbiography}
\begin{IEEEbiography}[{\includegraphics[width=1in,height=1.25in,clip,keepaspectratio]{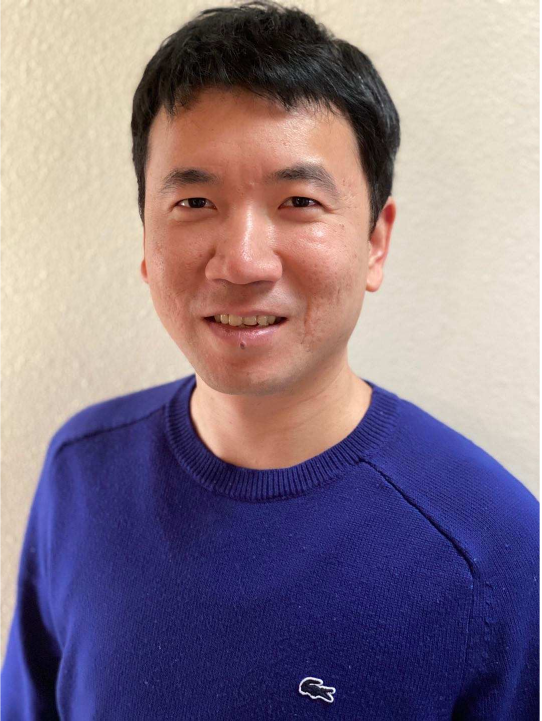}}]{Xiang Sun} [S'13, M'18] is an Assistant Professor with the Department of Electrical and Computer Engineering at the University of New Mexico. He received his Ph.D. degree in Electrical Engineering from New Jersey Institute of Technology (NJIT) in 2018, and his M.E. and B.E. degrees both from Hebei University of Engineering in 2011 and 2008, respectively. His research interests include free space optics, wireless networks, distributed/cooperative machine learning, Internet of Things, edge computing, and green communications and computing. He has received several honors and awards, including NJIT Ross Fellowship 2014-2015, 2016 IEEE International Conference on Communications (ICC) Best Paper Award, 2017 IEEE Communications Letters Exemplary Reviewers Award, 2018 NJIT Hashimoto Price, 2018 InterDigital Innovation Award on IoT Semantic Mashup, and 2019 IEICE Communications Society Best Tutorial Paper Award. He is an Associate Editor of the IEEE Open Journal of the Computer Society and Elsevier Digital Communications and Networks.
	\end{IEEEbiography}
	
\begin{IEEEbiography}{Rana Albelaihi}[S'20] 
	received her M.S. degree in computer science from Tennessee State University, Tennessee, USA in 2016. She received her Phd in computer science at the University of New Mexico in 2022. Her research interests include Internet of Things, federated learning, and green communications and computing.
\end{IEEEbiography}

\begin{IEEEbiography}[{\includegraphics[width=1in,height=1.25in,clip,keepaspectratio]{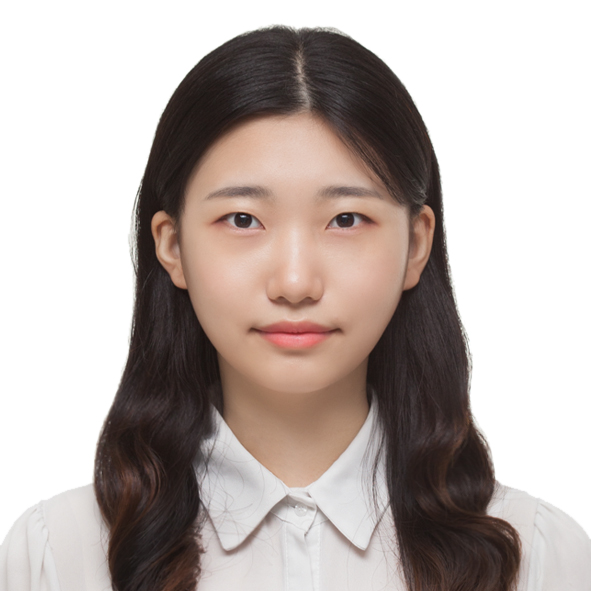}}]{Chaeeun Park}[S'23] obtained her B.S in Computer Science from University of New Mexico in 2022. She started her academic research at the SENet lab at the University of New Mexico in 2022 and commenced her master's program in 2023. Specializing in machine learning and wireless networks, her research explores various topics, including clustering federated learning and drone-assisted mobile access networks.
\end{IEEEbiography}

 \begin{IEEEbiography}[{\includegraphics[width=1in,height=1.25in,clip,keepaspectratio]{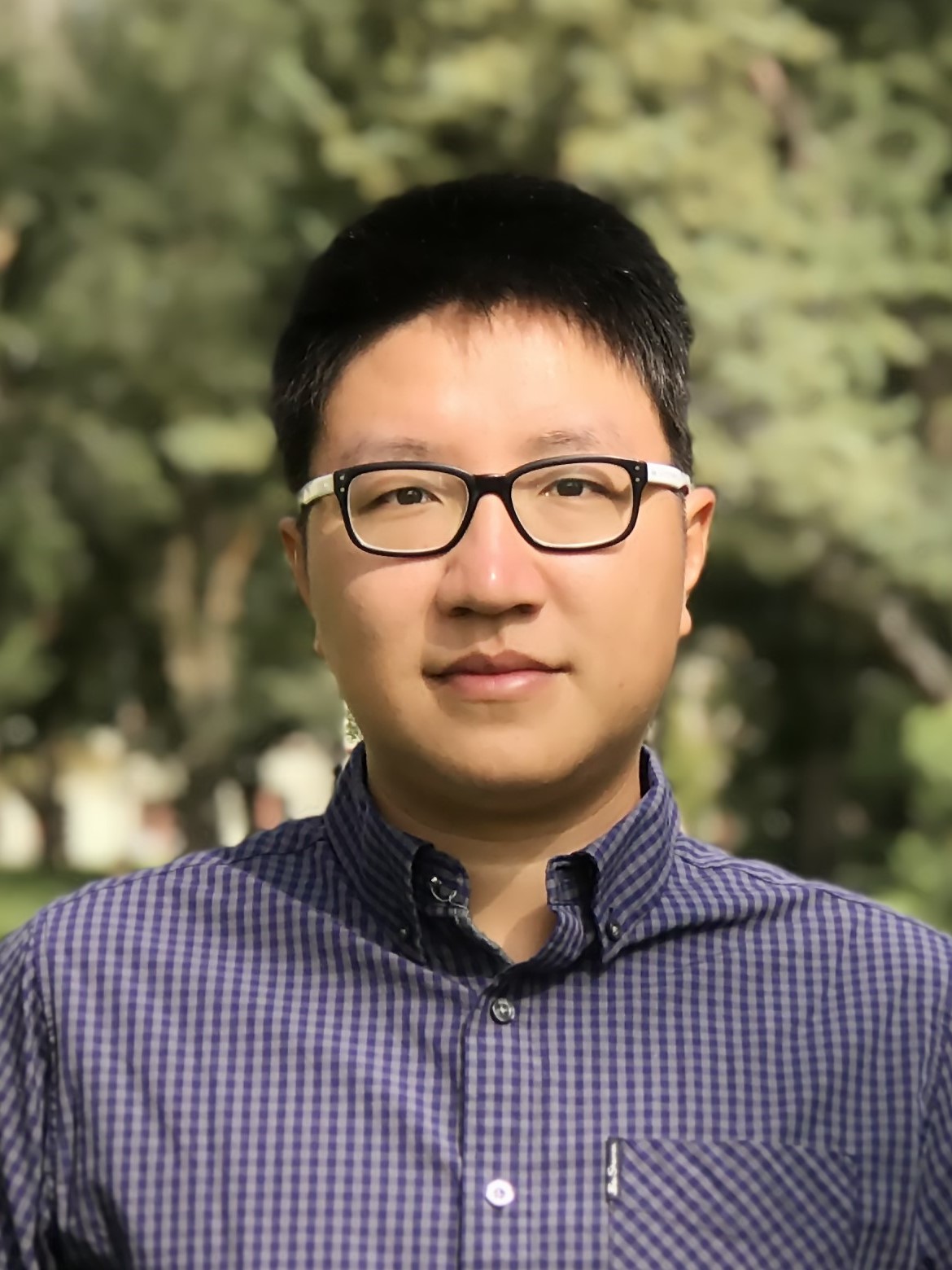}}]{Sihua Shao} [M'18, SM'23] received the Ph.D. degree in Electrical Engineering, from New Jersey Institute of Technology, Newark, NJ, USA, in 2018. He currently serves as an Assistant Professor with the Department of Electrical Engineering, New Mexico Tech, Socorro, NM, USA. He is also a Faculty Hire with the New Mexico SMART Grid Center. His area of expertise is wireless communication and networking, mainly focusing on reconfigurable intelligent surfaces, integrated sensing and communication, optical wireless communication, backscatter communication, machine learning in communications and networking, wireless communication and networking in harsh environments, UAV assisted wireless networks, heterogeneous radio-optical wireless networks, positioning, navigation and timing. Dr. Shao received the Hashimoto Prize for best doctoral dissertation from New Jersey Institute of Technology in 2018.
\end{IEEEbiography}

\end{document}